\documentclass[apj]{emulateapj}

\usepackage{natbib}
\usepackage{color}
\usepackage{comment}

\newcommand{\rev}{\textcolor{black}}

\accepted{to ApJ, January 6, 2016}

\begin{document}

\title{Demonstrating High-precision, Multi-band Transit Photometry with MuSCAT: A Case for HAT-P-14b}

\author{Akihiko~Fukui\altaffilmark{1}, Norio~Narita\altaffilmark{2,3,4}, Yui~Kawashima\altaffilmark{5},  Nobuhiko~Kusakabe\altaffilmark{2,3}, Masahiro~Onitsuka\altaffilmark{3,4},  Tsuguru~Ryu\altaffilmark{3,4}, Masahiro~Ikoma\altaffilmark{5}, Kenshi~Yanagisawa\altaffilmark{1}, and Hideyuki~Izumiura\altaffilmark{1,4}}
\email{afukui@oao.nao.ac.jp}

\altaffiltext{1}{Okayama Astrophysical Observatory, National Astronomical Observatory of Japan, Asakuchi, Okayama 719-0232, Japan}
\altaffiltext{2}{Astrobiology Center, 2-21-1 Osawa, Mitaka, Tokyo, 181-8588, Japan}
\altaffiltext{3}{National Astronomical Observatory of Japan, 2-21-1 Osawa, Mitaka, Tokyo 181-8588, Japan}
\altaffiltext{4}{SOKENDAI (The Graduate University for Advanced Studies), 2-21-1 Osawa, Mitaka, Tokyo 181-8588, Japan}
\altaffiltext{5}{Department of Earth and Planetary Science, Graduate School of Science, The University of Tokyo, 7-3-1 Bunkyo-ku, Tokyo 113-0033, Japan}

\begin{abstract}
The Multicolor Simultaneous Camera for studying Atmospheres of Transiting exoplanets (MuSCAT) is an optical three-band ($g'_2$-, $r'_2$-, and $z_\mathrm{s,2}$-band) imager that was recently developed for the 188~cm telescope at Okayama Astrophysical Observatory with the aim of validating and characterizing transiting planets. 
In a pilot observation with MuSCAT we observed a primary transit of HAT-P-14b, a high-surface gravity ($g_\mathrm{p}=38$~ms$^{-2}$) hot Jupiter around a bright ($V=10$) F-type star. From a 2.9~hr observation we achieved the five-minute binned photometric precisions of 0.028\%, 0.022\%, and 0.024\% in the $g'_2$, $r'_2$, and $z_\mathrm{s,2}$ bands, respectively, which proovided the highest-quality photometric data for this planet. 
Combining these results with those of previous observations, we search for variations of transit timing and duration over five years as well as variations of planet-star radius ratio ($R_\mathrm{p}/R_\mathrm{s}$) with wavelengths, but can find no considerable variation in any parameters.
On the other hand, using the transit-subtracted light curves we simulate achievable measurement error of $R_\mathrm{p}/R_\mathrm{s}$ with MuSCAT for various planetary sizes, assuming three types of host stars: HAT-P-14, the nearby K-dwarf HAT-P-11, and the nearby M-dwarf GJ1214.
Comparing our results with the expected atmospheric scale heights, we find that MuSCAT is capable of probing the atmospheres of planets as small as a sub-Jupiter ($R_\mathrm{p} \sim 6 R_\oplus$) around HAT-P-14 in all bands, a Neptune ($\sim4R_\oplus$) around HAT-P-11 in all bands, and a super-Earth ($\sim2.5R_\oplus$) around GJ1214 in $r'_2$ and $z_\mathrm{s,2}$ bands. 
These results promise that MuSCAT will produce fruitful scientific outcomes in the K2 and TESS era.
\end{abstract}


\section{Introduction}

Ground-based multiband photometry is an important tool for validating and characterizing transiting planets. Both ground- and space-based transit surveys often suffer from false positive detections caused by the contamination of eclipsing binaries, which can be identified by wavelength-dependent variations of transit depths \citep{2004A&A...425.1125T,2006ApJ...644.1237O,2011PASP..123.1391C}. The wavelength dependence of transit depth can also be used to characterize the atmospheres of confirmed transiting planets. Although the multiband photometry generally does not have the power to resolve specific atoms or molecules, it is still useful to observe broad spectral features and test for the presence or absence of cloud/haze particles \citep[e.g.,][]{2011ApJ...736...78C,2012A&A...538A..46D,2013ApJ...771..109D,2013ApJ...770...95F,2014ApJ...790..108F,2013MNRAS.436....2M,2014A&A...562A.126M,2013PASJ...65...27N,2013ApJ...773..144N,2013A&A...559A..32N,2015A&A...579A.113N}.

Simultaneity of the multiband photometry is also important. The transit depth can vary with time when the host star is spotted, which can be a source of systematics in the atmospheric study \citep[e.g.,][]{2008MNRAS.385..109P}. This problem can be reduced by simultaneous multiband observations. Furthermore, the multiple observations for a single transit will provide accurate measurements of the transit parameters related to the planetary orbit at the moment, allowing us to search for the transit timing variations (TTVs) and transit duration variations (TDVs) with a good precision. In particular, the ongoing and planned space-based transit surveys such as K2 \citep{2014PASP..126..398H} and TESS \citep{2015JATIS...1a4003R}, have limited survey periods for a particular survey field, making the ground-based, high-precision photometric followups essential to measure the planetary mass via TTVs \citep{2015MNRAS.454.4267B,2015ApJ...815...47N}.

A multiband simultaneous imager is thus useful for the above sciences. 
Telescopes with a 1--2~m aperture size are optimal for such instruments in the sense that a high enough photometric precision for the above sciences can be achieved and a relatively large amount of telescope time can be available compared with larger telescopes. However, currently only a handful of such instruments exist worldwide \citep[e.g., GROND, BUSCA, and SIRIUS,][]{2009A&A...497..545S,2010A&A...510A.107M,2013PASJ...65...27N}.

The Multicolor Simultaneous Camera for studying Atmospheres of Transiting exoplanets (MuSCAT) is a multiband imager that was recently developed for the 188~cm telescope at Okayama Astrophysical Observatory (OAO) in Japan with the aim of studying transiting exoplanets \citep{2015JATIS...1d5001N}. 
MuSCAT can simultaneously obtain three-band images through the SDSS $g'_2$-, $r'_2$-, and $z_\mathrm{s,2}$-band filters, each using a 1024 $\times$ 1024 pixels CCD with the field of view (FOV)\footnote{The FOV is extensible to the maximum of $\sim$12$'$ $\times$ 12$'$ with larger-format CCDs.} of 6\farcm1 $\times$ 6\farcm1 . This is the first high-precision multiband imager for a 2~m class telescope in the East Asia region (at $\sim$133\fdg5 in longitude), filling the in blank for the ground-based transit-followup networks \citep[e.g., the KELT followup network,][]{2012ApJ...761..123S}.

In this paper we report the results of a pilot observation of MuSCAT, which targeted a primary transit of HAT-P-14b (aka WASP-27b), a hot Jupiter orbiting a bright ($V$=10.1) F-type star in a retrograde and slightly eccentric ($e=0.1$) orbit with the period of 4.63 days \citep{2010ApJ...715..458T,2011AJ....141..161S,2011AJ....141...63W}. 
So far only one photometric followup observation has been reported after the discoveries \citep{2011AA...527A..85N}, leaving room for further photometric followups to search for such as TTVs, TDVs, and transit-depth variations. The main reason for selecting this system for a pilot observation is that a good comparison star with a similar brightness to the host star  ($V$=9.6) exists within the FOV (HIP 84832; separated from HAT-P-14 by 4.4$'$), offering a good opportunity to  achieve a high photometric precision. Stars of similar brightness do not always exist in the same FOV as such a bright target star. Because one of the main targets of MuSCAT will be super-Earths and Neptunes around bright stars ($V<10$) that are expected to be discovered by such as K2 and TESS, it is worthwhile to investigate how photometric precision can be achieved with MuSCAT for a $V$=10 star. Although \citet{2015JATIS...1d5001N} already demonstrated a high-precision photometry for a $V$=10 star with MuSCAT, it was from an only one-hour-long out-of-transit (OOT) observation.  This is the first such demonstration from a full-transit observation.

The rest of this paper is organized as follows. Sections \ref{sec:observation} and \ref{sec:analysis} describe the observation and analysis, respectively. Section \ref{sec:results} shows the results and discussions, including a simulation of the achievable measurement error of the planet-to-star radius ratio with MuSCAT for various sizes of planets around three types of host stars. Finally we summarize our work in Section \ref{sec:summary}.

\section{Observation}
\label{sec:observation}
\begin{figure*}
\begin{center}
\vspace{10pt}
\includegraphics[width=17cm]{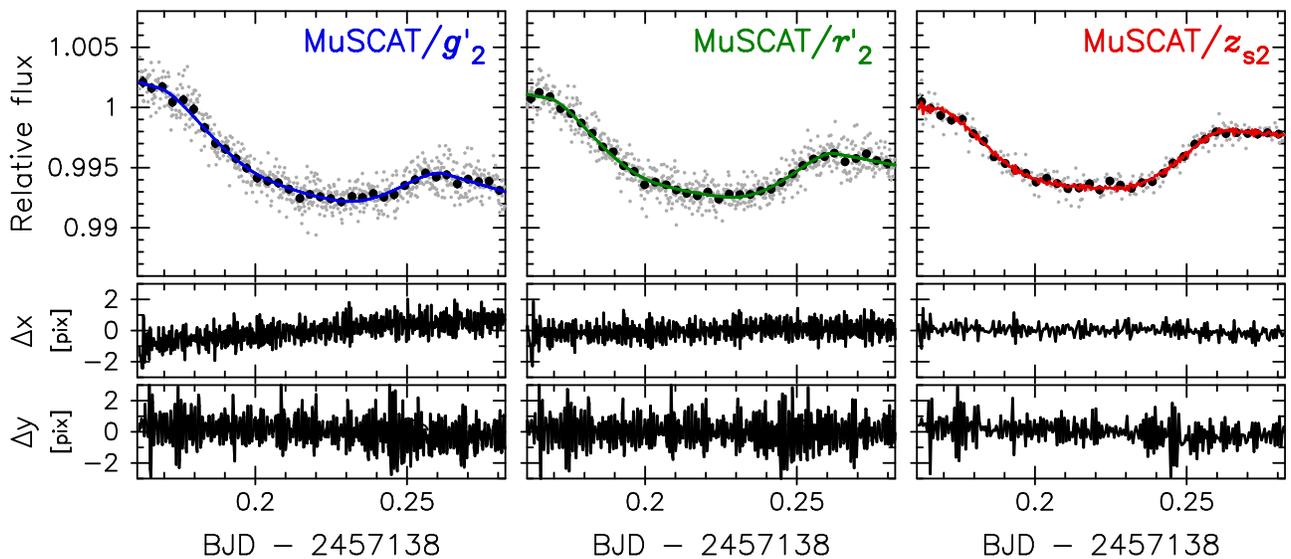}
\vspace{5pt}
\caption{
(Top row) The relative light curves of HAT-P-14 observed on 2015 April 25 using the OAO 188~cm telescope/MuSCAT. The left, middle, and right panels are for $g'_2$, $r'_2$, and $z_\mathrm{s,2}$ bands, respectively. The gray and black circles represent the unbinned data and five minutes binned data, respectively. The solid lines are the best-fit OOT+transit models derived in Section \ref{sec:fit_muscat}. (Middle row) The relative centroid positions of stars on the CCDs in the $X$ direction. (Bottom row) The same as the middle row panels, but in the $Y$ direction.
\label{fig:uncorrected}
}
\end{center}
\end{figure*}

We conducted a photometric observation for a primary transit of HAT-P-14b by using the OAO 188~cm telescope equipped with MuSCAT on 2015 April 25 UT. The coordinate of the FOV center was selected such that HAT-P-14 and two comparison stars of HIP~84832 and TYC~3086-78-1 ($V$=10.7) were imaged together. We started the observation $\sim$10 minutes before the transit and continued it for 2.9~hr covering the full 2.2~hr transit. The sky was perfectly clear with no moon visible during the observation.  The $g'_2$- and $r'_2$-band images were taken with the exposure time of 10 s in high-speed (2 MHz) readout mode,  while the $z_\mathrm{s,2}$ band images were taken with the exposure time of 30 s in low-seed (100 kHz) readout mode\footnote{The reason for using the low-speed readout mode was that the CCD camera for $z_\mathrm{s,2}$ band produced uncorrectable systematic noises only in the high-speed readout mode. The camera was repaired after the observation and the problem has been fixed.}, resulting in the observing cadences of 14 s, 14 s, and 43 s for the $g'_2$, $r'_2$, and $z_\mathrm{s,2}$ bands, respectively. The exposure time and the total number of data points for each band are listed in Table \ref{tbl:data_list}. During the observation we activated the self autoguiding system \citep{2015JATIS...1d5001N} using the $r'_2$-band channel so as to keep the stellar positions on the detectors within $\sim$1 pixel. The measured displacements of the stellar centroids during the observation are listed in Table \ref{tbl:lc_muscat} and are shown in Figure \ref{fig:uncorrected}.

\tabletypesize{\small}
\begin{deluxetable*}{cccccccc}
\tablecaption{List of Light Curves Analyzed in This Work
 \label{tbl:data_list}
}
\tablefontsize{\small}
\tablehead{
Obs. Date & Telescope & Filter & Exposure &  $N_\mathrm{data}$ & rms & $\beta$ & References\\
(UT) & & & (s) & & (\%) &\\
}
\startdata
2015 Apr 25 & OAO 188~cm & Sloan $g'_2$ & 10 & 748 & 0.103 & 1.27 & This work\\
2015 Apr 25 & OAO 188~cm & Sloan $r'_2$ & 10 & 754 & 0.091 & 1.00 & This work\\
2015 Apr 25 & OAO 188~cm & Sloan $z_\mathrm{s,2}$ & 30 &  242 & 0.068 & 1.00 & This work\\
2009 Mar 25 & FLWO 1.2~m &  Sloan $i'$ & 20 & 470 & 0.171 & 1.00 & \cite{2010ApJ...715..458T}\\
2009 Apr 8 & FLWO 1.2~m & Sloan $i'$ & 15  & 359 & 0.191 & 1.06 &\cite{2010ApJ...715..458T}\\
2009 May 15 & FLWO 1.2~m  & Sloan $i'$ & 15 & 304 & 0.129 & 1.17 &\cite{2010ApJ...715..458T}\\
2010 Mar 12 & Asiago 1.82~m & $R$ & 120$^a$ & 104 & 0.056 & 1.03 & \cite{2011AA...527A..85N}\\
2010 May 21 & FTN & Pan-STARRS $Z$  & 20 & 343 & 0.111 & 1.09 & \cite{2011AJ....141..161S}\\
2010 Jun 22 & LT & $V$ & 30$^b$ & 281 & 0.072 & 1.18 & \cite{2011AJ....141..161S}
\enddata
\tablecomments{
\tablenotetext{a}{\ Original exposure time was 2 s. The 120 s binned data are publicly available.}
\tablenotetext{b}{\ Original exposure time was 3.7 s.  The 30 s binned data are provided by E.~K.~Simpson et al. (private communication).}
}
\end{deluxetable*}
\begin{deluxetable*}{lccccccc}
\tablecaption{Transit Light Curves of HAT-P-14b Obtained with MuSCAT
\label{tbl:lc_muscat}}
\tablehead{
Filter & Time & Flux & OOT-corrected & Error \tablenotemark{a} & Air Mass & $\Delta x$ & $\Delta y$\\
& (BJD$_\mathrm{TDB}$ - 2,450,000) & & Flux & & & (Pixels) & (Pixels)  
}
\startdata
$g'_2$ & 7138.161833 & 1.00754 & 0.99946 & 0.00152 & 1.1367 & -1.11 & 0.06\\
$g'_2$ & 7138.161972 & 1.01017 & 1.00207 & 0.00152 & 1.1363 & -1.60 & 0.52\\
$g'_2$ & 7138.162215 & 1.00899 & 1.00092 & 0.00152 & 1.1356 & -1.62 & 0.27\\
$g'_2$ & 7138.162366 & 1.00849 & 1.00044 & 0.00151 & 1.1351 & -1.87 & 0.52\\
$g'_2$ & 7138.162505 & 1.00793 & 0.99989 & 0.00151 & 1.1347 & -2.04 & 0.83\\
... &&&&&&&
\enddata
\tablecomments{
\\
$^a$\ Error is rescaled by the red noise factor $\beta$.\\
(This table is available in its entirety in machine-readable form in the version published online.)
}
\end{deluxetable*}

\section{Analysis}
\label{sec:analysis}

\subsection{Data Reduction}

All of the observed images are corrected for dark and flat fields in a standard manner. 
The flat-fielding frame for each band is created from one hundred dome-flat images taken on the observing night. Then aperture photometry is performed for the target star (HAT-P-14) and two comparison stars by a customized tool with a constant aperture radius mode \citep{2011PASJ...63..287F}. 
After several trials with different aperture radii, we select the optimal aperture radius such that the resultant relative light curve produced by dividing the target-star flux by the sum of the fluxes of the two comparison stars gives the minimum dispersion relative to the best-fitted transit model. The selected aperture radii are 26, 27, and 24 pixels for the $g'_2$, $r'_2$, and $z_\mathrm{s,2}$ bands, respectively.
The time for each data point is assigned as the center of exposure in the Barycentric Julian Day (BJD) time system based on Barycentric Dynamical Time (TDB), which is converted from Coordinated Universal Time (UTC) recorded on the FITS header via the code of \citet{2010PASP..122..935E}. The produced relative light curves are listed in Table \ref{tbl:lc_muscat} and are shown in Figure \ref{fig:uncorrected}.

\subsection{Light Curve Fitting to the MuSCAT Light Curves}
\label{sec:fit_muscat}

To compare the quality of the observed light curves with the results of previous work we fit the MuSCAT light curves with transit models.
First we determine the OOT function for each light curve by fitting the entire light curve with transit+OOT models. A transit+OOT model is expressed as
\begin{eqnarray}
\label{eq:OOT1}
F &=&  \rev{\left( k_0 + \sum_{i=1} k_i X_i \right)} \times F_\mathrm{tr},
\end{eqnarray}
where $F$ is the relative flux, $F_\mathrm{tr}$ is the transit light curve model, $\{{\bf X}\}$ are variables, and $\{{\bf k}\}$ are the coefficients. For $F_\mathrm{tr}$, we use the analytic formula given by \citet{2009ApJ...690....1O},  which is equivalent to that of \citet{2002ApJ...580L.171M} when using the quadratic limb-darkening raw. The transit model includes the following nine parameters: the mid-transit time, $T_\mathrm{c}$; the orbital period, $P$; the planet-to-star radius ratio, $R_\mathrm{p}/R_\mathrm{s}$; the semi-major axis in units of stellar radius, $a/R_\mathrm{s}$; the minimum sky-projected planet-star distance in units of stellar radius (impact parameter), $b$; the eccentricity, $e$; the longitude of periastron, $\omega$; and the quadratic limb-darkening coefficients, $u_1$ and $u_2$. 
For $\{{\bf X}\}$, we try different combinations among the following variables: the time, $t$; the square of time, $t^2$; the airmass, $z$; and stellar displacement on the CCD, $\Delta x$ and $\Delta y$. The optimal combination is determined such that the Bayesian information criteria, BIC ($\equiv \chi^2 + k \ln N$) \citep{1978Schwarz}, for the best-fit model is minimum. The $\chi^2$ value is given by 
\begin{eqnarray}
\label{eq:chi2}
\chi^2 = \sum_i \left( \frac{f_{\mathrm{obs},i}-f_{\mathrm{model},i}}{\sigma_i} \right)^2,
\end{eqnarray}
where $f_{\mathrm{obs},i}$ and $f_{\mathrm{model},i}$ are the $i$-th measured and model fluxes, respectively, and $\sigma_i$ is the $i$-th flux error.
The fit is done by using the AMOEBA algorithm \citep{1992nrca.book.....P}, letting $T_\mathrm{c}$, $a/R_\mathrm{s}$, $b$, $R_\mathrm{p}/R_\mathrm{s}$, and $\{{\bf k}\}$ be free while fixing $P$, $e$, and $\omega$ at 4.627682 d \citep{2011AA...527A..85N}, 0.115, and  98.8$^\circ$ \citep{2014ApJ...785..126K}, respectively.
We also fix $u_1$ and $u_2$ at the theoretical values of \{$u_1$, $u_2$\} = \{0.440, 0.287\}, \{0.289, 0.322\}, \{0.304, 0.166\} for the $g'_2$, $r'_2$, and $z_\mathrm{s,2}$ bands, respectively.
These are the values for a star with the effective temperature of $T_\mathrm{eff}=6500$K, the surface gravity of $\log g$=4.0, the metallicity of $\log \mathrm{[M/H]}$=0, and the micro-turbulent velocity of $\xi$=\rev{1}~km~s$^{-1}$, calculated for the standard SDSS $g'$-, $r'$-, and $z'$-band filters by \citet{2011A&A...529A..75C}.
As a result, we determine $\{{\bf X}\}$ = $\{t\}$, $\{t\}$, and $\{t, \Delta x\}$ for $g'_2$, $r'_2$, and $z_\mathrm{s,2}$ bands, respectively.
We note that the OOT model selection using BIC does not guarantee that the selected model is robust unless the BIC value is significant over the other unselected models and therefore the choice of the OOT model could affect the final transit parameters. Among the unselected OOT models, some indeed have similar BIC values to the selected model within 5. 
However, when we apply the unselected similar-BIC OOT models, we find that the derived values of $a/R_\mathrm{s}$, $b$, and $R_\mathrm{p}/R_\mathrm{s}$  become either to be very close to those derived using the least-BIC models or have too large variations among the three bands, implying
that our choice of the OOT model is robust and the possible bias on the transit parameters should be small. We therefore neglect this possible bias in this paper. Note that more sophisticated methods that properly estimate uncertainties associated with the choice of the OOT model have been proposed  \citep[e.g.][]{2014MNRAS.445.3401G}.

Next, for each light curve we rescale the initial flux uncertainties such that the reduced $\chi^2$ of the transit+OOT model fit becomes unity. In addition we further rescale them by a factor of $\beta$ \citep{2008ApJ...683.1076W,2006MNRAS.373..231P}, taking time-correlated (red) noises into account where $\beta$ is the ratio of the actual standard deviation of a binned residual light curve to the one expected from the unbinned residual light curve assuming white noises. We take a mean value of $\beta$ for the binning size of between 5 and 10 minutes. The applied $\beta$ values are summarized in Table \ref{tbl:data_list}. 
We note that this $\beta$-scaling method models the red noises in only an approximation way and that more sophisticated methods that directly treat time correlation of the noises, such as Gaussian processes \citep{2012MNRAS.419.2683G}, have been proposed.

Finally, to properly derive the values and uncertainties of the parameters we perform the Markov Chain Monte Carlo (MCMC) analysis using a customized code \citep{2007PASJ...59..763N,2013PASJ...65...27N}. The light curves are fitted with the transit+OOT models with the OOT functions selected above. 
We first analyze each of the $g'_2$-, $r'_2$-, and $z_\mathrm{s,2}$-band light curves independently and then analyze all of the three light curves jointly.
In an experimental fitting we found that $u_1$ and $u_2$ cannot meaningfully be constrained from the light curves alone because $u_1$ and $u_2$ are strongly correlated with each other, and the transit impact parameter $b$ is close to unity (0.91) so that the planet transits across the limb of the star where the intensity does not vary so much. Therefore, in this MCMC analysis we use the alternative parameterization of $w_1 \equiv u_1 \cos \phi - u_2 \sin \phi$ and $w_2 \equiv u_1 \sin \phi + u_2 \cos \phi$ with $\phi = 40^\circ$, which are much less correlated \citep{2008MNRAS.390..281P,2011A&A...529A..75C}. 
Then we let both $w_1$ and $w_2$ be free in the MCMC process while imposing priors on them in different ways. For $w_2$, 
we use the following $\chi^2$ merit function
assuming a Gaussian prior:
\begin{eqnarray}
\label{eq:merit}
\chi'\ ^2 = \chi^2 + \sum_{i=1} \left( \frac{p_i - p_{i,\mathrm{prior}}}{\sigma_{p_{i,\mathrm{prior}}}} \right)^2,
\end{eqnarray}
where $\chi^2$ is given by Equation (\ref{eq:chi2}), $p_i$ is a parameter value (here, $w_2$), and $p_{i,\mathrm{prior}}$ and $\sigma_{p_{i,\mathrm{prior}}}$ are the prior value and its 1-$\sigma$ uncertainty, respectively.
As the prior value, we adopt the theoretical value calculated from \citet{2013A&A...552A..16C}. Regarding the uncertainty of the prior, we find that the theoretical value of $w_2$ does not vary beyond $\pm 0.01$ within the allowed stellar parameters (temperature, surface gravity, metallicity, and microtarbulance velocity),  regardless of whether the stellar model is based on ATLAS \citep{2011A&A...529A..75C} or PHOENIX \citep{2013A&A...552A..16C} and the computation method is least square or flux conservation \citep[see, ][]{2009A&A...506.1335C}. Nevertheless, taking into account the possible systematics in the theoretical models and the imperfect match of the filter transmission functions between the theoretical models and observations, we conservatively adopt 0.05 as $\sigma_{w_2,\mathrm{prior}}$ for all bands. On the other hand, because $w_1$ 
has large variations (up to $\sim$0.2) in the theoretical values within the allowed stellar parameters, the based stellar models, and the computation methods, we only set upper and lower limits on $w_1$ among these possible theoretical values. We summarize the prior values of $w_1$ and $w_2$ in Table \ref{tbl:w1w2}.
Regarding the other parameters, we allow $T_\mathrm{c}$, $a/R_\mathrm{s}$, $b$, $R_\mathrm{p}/R_\mathrm{s}$, $e \sin \omega$, $e \cos \omega$, and $\{{\bf k}\}$ to vary freely while imposing priors on $e \sin \omega$ and $e \cos \omega$ to the values of \citep{2014ApJ...785..126K} using the $\chi^2$ merit function of Equation (\ref{eq:merit}). In the joint analysis, we treat $w_1$, $w_2$, and $\{{\bf k}\}$ as independent parameters for each band, while commonly treat the other parameters.
In each MCMC run, the starting values of the parameters are set to the best-fit values determined by the AMOEBA algorithm. We perform 10 independent MCMC chains with 10$^6$ steps each and then merge all the steps except for the first 2$\times10^5$ steps as burn-in to construct a posterior probability distribution for each parameter. The widths of the Gaussian jump functions are adjusted such that the success ratio of the MCMC runs becomes $\sim$20-30\%. We confirm the convergence of the MCMC runs by checking the consistency of the results from the 10 independent chains.

\begin{deluxetable*}{lccccc}
\tablecaption{\rev{Priors and MCMC Results of $w_1$ and $w_2$} \label{tbl:w1w2}}
\tablehead{
Telescope & Filter & $w_\mathrm{1,prior}$ & $w_\mathrm{2,prior}$ & $w_\mathrm{1,mcmc}$ & $w_\mathrm{2,mcmc}$
}
\startdata
OAO 188~cm & $g'_2$ & [0.120,  0.302] &  0.505 $\pm$ 0.050 & 0.217 $^{+0.059}_{-0.065}$ & 0.525 $\pm$ 0.043\\
OAO 188~cm & $r'_2$ & [-0.012, 0.207] & 0.424 $\pm$ 0.050 & 0.102 $^{+0.072}_{-0.077}$ & 0.433 $\pm$ 0.042\\
OAO 188~cm & $z_\mathrm{s,2}$ & [-0.042, 0.149] & 0.320 $\pm$ 0.050 & 0.052 $\pm$ 0.065 & 0.327 $\pm$ 0.045\\
FLWO 1.2~m & $i'$ & [-0.061, 0.159] & 0.363 $\pm$ 0.050 & 0.040 $^{+0.079}_{-0.070}$ & 0.344 $\pm$ 0.049\\
Asiago 1.82~m & $R$ & [-0.027, 0.204] & 0.409 $\pm$ 0.050 & 0.096 $^{+0.075}_{-0.082}$ & 0.426 $\pm$ 0.051\\
FTN & $Z$ & [-0.042, 0.149] & 0.320 $\pm$ 0.050 & 0.058 $^{+0.063}_{-0.067}$ & 0.332 $\pm$ 0.049\\
LT & $V$ & [0.009, 0.225] & 0.436 $\pm$ 0.050 & 0.114 $\pm$ 0.074 & 0.434 $\pm$ 0.049
\enddata
\end{deluxetable*}

\begin{figure*}
\begin{center}
\vspace{10pt}
\includegraphics[width=17cm]{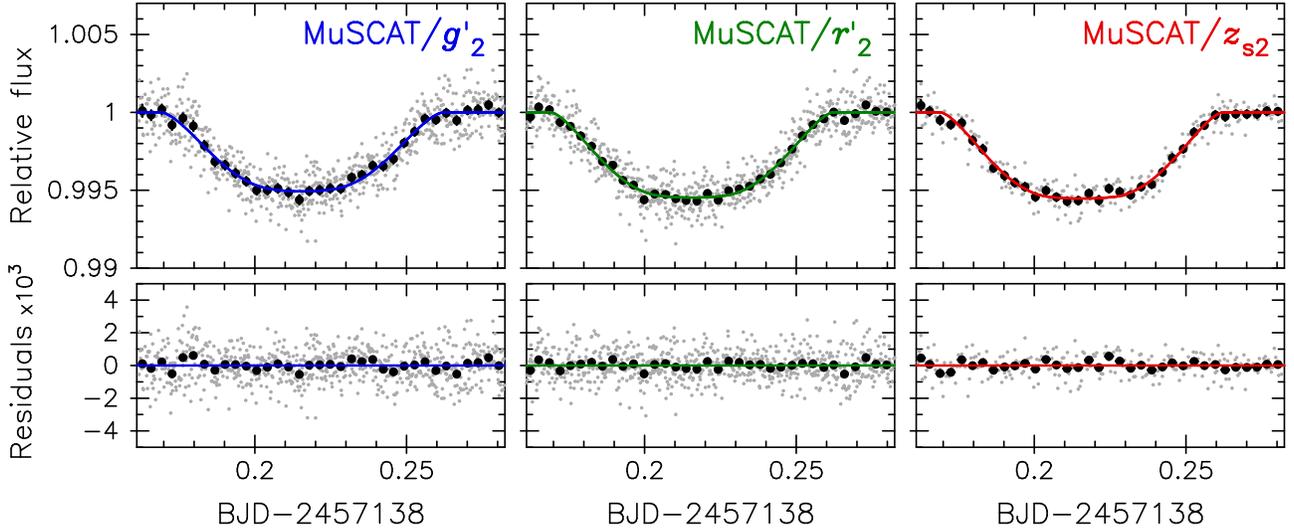}
\vspace{5pt}
\caption{
(Top row) The OOT-corrected light curves of MuSCAT. The left, middle, and right panels are for $g'_2$, $r'_2$, and $z_\mathrm{s,2}$ bands, respectively. The meanings of the gray and black circles are the same as in Figure \ref{fig:uncorrected}. The solid lines are the best-fit transit models derived in Section \ref{sec:fit_muscat}. (Bottom row) The same as the top-row panels, but residual light curves.
\label{fig:corrected}
}
\end{center}
\end{figure*}
\tabletypesize{\footnotesize}
\begin{deluxetable}{lccc}
\tablecaption{The MCMC Results for the MuSCAT Data\label{tbl:fit_muscat}}
\tablewidth{8.5cm}
\tablehead{
& $a/R_\mathrm{s}$ & $b$ & $R_\mathrm{p}/R_\mathrm{s}$
}
\startdata
MuSCAT ($g'_2$) & 7.66 $^{+0.67}_{-0.95}$ & 0.935 $^{+0.124}_{-0.025}$ & 0.0900 $^{+0.0843}_{-0.0083}$\\
MuSCAT ($r'_2$) & 8.30 $\pm$ 0.28  & 0.908 $\pm$ 0.008 & 0.0821 $^{+0.0018}_{-0.0016}$ \\ 
MuSCAT ($z_\mathrm{s,2}$) & 8.56 $\pm$ 0.39 & 0.903 $\pm$ 0.009 & 0.0777 $^{+0.0018}_{-0.0016}$ \\ 
MuSCAT (all) & 8.36 $\pm$ 0.23 & 0.9072 $^{+0.0057}_{-0.0051}$ &  0.0805 $\pm$ 0.0011\\
\citet{2012MNRAS.426.1291S} & 8.26 $\pm$ 0.25 & --- &  0.0831 $\pm$ 0.0022
\enddata
\end{deluxetable}

\subsection{Homogeneous Light Curve Analysis Including Published Data}
\label{sec:fit_all}

To search for possible variations of transit parameters we re-analyze both the MuSCAT data and published ones simultaneously. From literature we use six full-transit light curves obtained by 1--2m class telescopes, which include three light curves obtained with the FWLO 1.2~m telescope ($i$-band) by \citet{2010ApJ...715..458T}, two with the 2.0~m FTN ($z$-band) and the 2.0~m LT ($V$-band) by \citet{2011AJ....141..161S} \footnote{These light curves are not publicly available, and we obtained them from the authors in private.}, and one with the Asiago 1.82~m telescope ($R$-band) by \citet{2011AA...527A..85N}. To equally treat all the light curves, we rescale the original uncertainties of the fluxes by the same procedure as done for the MuSCAT data, i.e., rescaling them so that the reduced $\chi^2$ for a transit model fit to each light curve becomes unity and then rescaling them again by the red noise factor $\beta$. 
The calculated $\beta$ factors are listed in Table \ref{tbl:data_list}.
Then we fit the total of nine light curves simultaneously by the same MCMC procedure as in Section \ref{sec:fit_muscat}, but this time letting $T_\mathrm{c}$,  $b$, $w_1$, and $w_2$ be free for each transit and $R_\mathrm{p}/R_\mathrm{s}$ be free for each band. Although the OOT trends of these published light curves had already been corrected in their own ways, we allow $\{{\bf X}\}$ = $\{t\}$ to vary freely for each light curve in order to propagate at least a part of the uncertainties of the OOT corrections into the final parameter uncertainties. Note that we cannot exactly reprocess their original OOT corrections, which might lead to some underestimation of the final uncertainties.
We also impose priors on $w_1$ and $w_2$ the same way as in Section \ref{sec:fit_muscat}. The applied prior values, as well as the derived values in this joint analysis, of $w_1$ and $w_2$ are listed in Table \ref{tbl:w1w2}.  This time we perform 10 independent MCMC chains with 10$^7$ steps each and merge all but the first 2$\times10^6$ (burn-in) steps to calculate the posterior probability distributions.

\begin{figure}
\begin{center}
\includegraphics[width=8.cm]{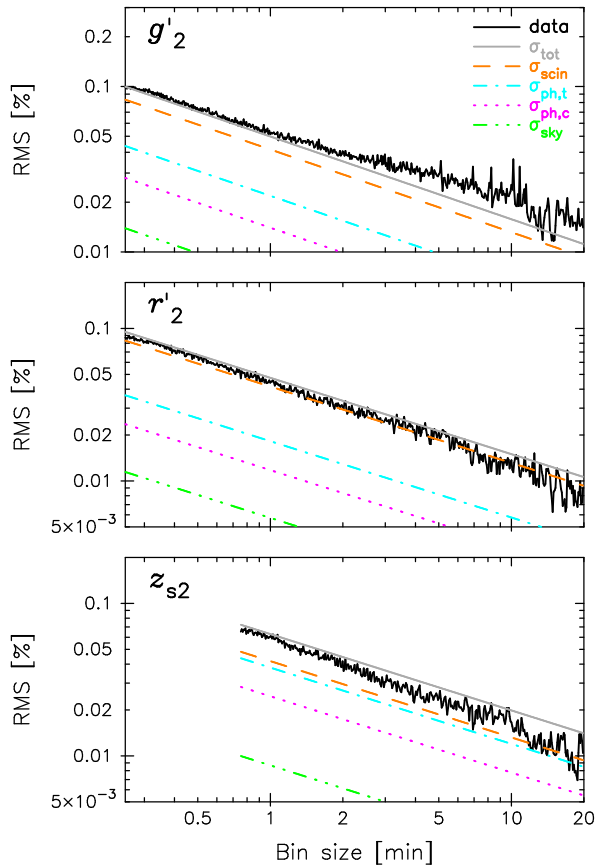}
\caption{Binned rms of the residual light curves of MuSCAT as a function of binning size. The top, middle, and bottom panels are for the $g'_2$-, $r'_2$-, and $z_{s,2}$-band light curves, respectively. In each panel the black line indicates the observed data while the orange dashed, light blue dash-dotted, magenta dotted, green three-dotted-dashed, and gray solid lines represent expected values from scintillation noise, photon noise of the target star, photon noises of the comparison stars, sky background noise, and the total of them, respectively. 
\label{fig:binned_rms}
}
\end{center}
\end{figure}

\section{Results and Discussions}
\label{sec:results}
\subsection{Quality of the MuSCAT Data}

In Table \ref{tbl:fit_muscat}, we list the measured values and their 1-$\sigma$ uncertainties of $a/R_\mathrm{s}$, $b$, and $R_\mathrm{p}/R_\mathrm{s}$ derived from the individual fit to the three MuSCAT light curves and from the joint fit to all of them (Section \ref{sec:fit_muscat}). The four values in each parameter are largely consistent with each other. To compare them with the results of previous work we also list the values from \citet{2012MNRAS.426.1291S} who derived them by analyzing nine published light curves that were obtained with 1--2 m class telescopes, including five with the FLWO 1.2~m telescope \citep{2010ApJ...715..458T}, two with the 2.0~m LT \citep{2011AJ....141..161S}, one with the 2.0~m FTN \citep{2011AJ....141..161S}, and one with the Asiago 1.82~m telescope \citep{2011AA...527A..85N}.  The values from the joint fit to all the MuSCAT light curves are consistent with those from \citet{2012MNRAS.426.1291S} within the uncertainties.  In addition the uncertainties from MuSCAT are smaller than those from \citet{2012MNRAS.426.1291S}, meaning that MuSCAT provides the highest-quality transit data of this planet from only a single transit observation.

The light curves of MuSCAT corrected by the best-fit OOT models and the residual light curves from the best-fit transit models are shown in Figure \ref{fig:corrected}.
The root mean square (rms) values of the unbinned (five minutes binned) residual light curves are 0.10\% (0.028\%), 0.091\% (0.022\%), and 0.068\% (0.024\%) for the $g'_2$, $r'_2$, and $z_\mathrm{s,2}$ bands, respectively. In Figure \ref{fig:binned_rms} we show the rms values of binned residual light curves as a function of binning size. 
The black lines indicate the observed data while the orange dashed, light blue dash-dotted, magenta dotted, green three-dotted-dashed, and gray solid lines represent expected values from scintillation noise, photon noise from the target star, photon noise from the comparison stars, sky background noise, and the total of them, respectively. The scintillation noise is calculated from the following \rev{approximation}:
\begin{eqnarray}
\label{eq:scin}
\sigma_\mathrm{scin} = 0.064 D^{-2/3} (\sec Z)^{7/4} e^{-h/h_0} T^{-1/2},
\end{eqnarray}
where $D$ (=188) is the diameter of the primary mirror of the telescope in cm, $Z$ is the zenith distance, $h$ (=372) is the height above sea level of the observatory in m, $h_0$ = 8000~m, and $T$ is the exposure time in seconds \citep{1967AJ.....72..747Y,1998PASP..110..610D}. 
Given that Equation (\ref{eq:scin}) is only a rough approximation, the observed photometric dispersions are largely consistent with the theoretical noise models. 
On the other hand, an excess above a slope extended from the binning size of $<$1~minutes is seen at the larger binning size in the $g'_2$ band, indicating that there exists red noise in the light curve as also indicated by the $\beta$ factor in Table \ref{tbl:data_list}. However, the level of this red noise is typical in ground-based photometric observations.

\subsection{Searching for TTVs, TDVs, and $R_\mathrm{p}/R_\mathrm{s}$ Variations with Wavelength}

\begin{figure}
\begin{center}
\includegraphics[width=8cm]{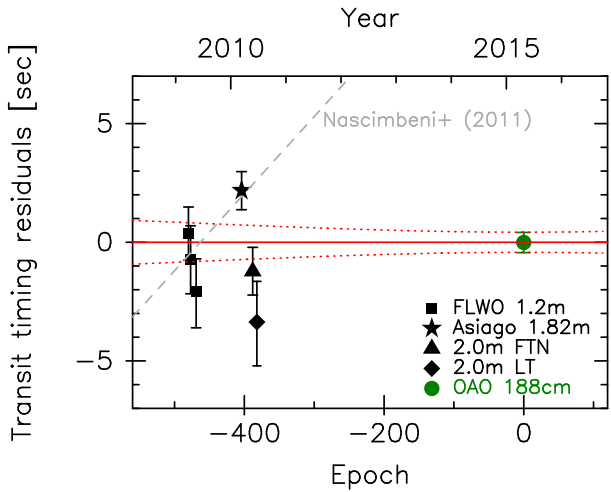}
\caption{Residuals of the measured transit timings ($T_\mathrm{c}$) from a linear ephemeris. The black squares, star, triangle, and diamond are the data from the FLWO 1.2~m telescope, the Asiago 1.82~m telescope, the 2.0~m FTN, and the 2.0~m LT, respectively, while the green circle is the data from the OAO 188~cm telescope with MuSCAT. The red dotted lines indicate the 1-$\sigma$ boundaries of the best-fit linear ephemeris. The gray dashed line is the ephemeris obtained by \citet{2011AA...527A..85N}.
\label{fig:TT}}
\end{center}
\end{figure}

From the results of the homogeneous analysis in Section \ref{sec:fit_all}, we first improve the transit ephemeris and search for TTVs. The measured $T_\mathrm{c}$ of the seven transits are listed in Table \ref{tbl:TT}. By fitting these data with a linear function we derive the improved transit ephemeris as
\begin{eqnarray}
T_\mathrm{c} \left(\mathrm{BJD}_\mathrm{TDB}\right) = \rev{7138.21545 \left(29 \right)} + \rev{4.6276594} \left(10 \right) \times E,
\end{eqnarray}
where $E$ is the relative transit epoch and the number in parentheses on the right represents the last two digits of 1$\sigma$ uncertainty. The residuals from this ephemeris, $\Delta T_\mathrm{c}$, are appended to Table \ref{tbl:TT} and are shown in Figure \ref{fig:TT}. The $\chi^2$ value of the linear fit is \rev{15.5} for five degrees of freedom (dof), meaning that the linear function nominally  has a 2.6~$\sigma$ discrepancy with the observed data. However, discrepancies with similar levels often arise in ground-based $T_\mathrm{c}$ observations, possibly due to unknown systematics rather than the true timing variations. Therefore, we do not consider it to be a noticeable TTV signal at this point. We note that the $T_\mathrm{c}$ value of the transit observed by \citet{2011AA...527A..85N} departs from our new ephemeris by \rev{2.8}~$\sigma$, causing a systematic shift in  their ephemeris (gray dashed line in Figure \ref{fig:TT}), which has propagated to be $\sim$15 minutes at the time of our observation. The reason for the departure is unknown, but in any case the correction of the ephemeris in this work should be useful for future observations. 

\tabletypesize{\small}
\begin{deluxetable*}{ccccc}
\tabletypesize{\footnotesize}
\tablecaption{The Measured Transit Timing ($T_\mathrm{c}$), Timing Residual from the Linear Ephemeris ($\Delta T_\mathrm{c}$), and Impact Parameter ($b$) for Each Transit\label{tbl:TT}}
\tablehead{
Transit Epoch & Telescope & $T_\mathrm{c}$ (BJD$_\mathrm{TDB}$ - 2450000) & $\Delta T_\mathrm{c}$ (days) & $b$  
}
\startdata
-480 & FLWO 1.2~m & 4916.93913 $\pm$ 0.00077 & 0.00022 & 0.9038 $\pm$ 0.0072 \\ 
-477 & FLWO 1.2~m & 4930.82139 $\pm$ 0.00099 & -0.00050 & 0.8972 $\pm$ 0.0091 \\
-469 & FLWO 1.2~m & 4967.84169 $\pm$ 0.00101 & -0.00148 & 0.8947 $\pm$ 0.0069 \\
-404 & Asiago 1.82~m & 5268.64254 $\pm$ 0.00054 & 0.00151 & 0.9022 $\pm$ 0.0061\\
-388 & FTN & 5342.68271 $\pm$ 0.00068 & -0.00088 & 0.9059 $\pm$ 0.0058 \\
-382 & LT & 5370.44714 $\pm$ 0.00125 & -0.00240 & 0.8854 $\pm$ 0.0077 \\
0 & OAO 188~cm & 7138.21544 $\pm$ 0.00029 & -0.00001 & 0.9062 $\pm$ 0.0042
\enddata
\end{deluxetable*}

\begin{figure}
\begin{center}
\includegraphics[width=8cm]{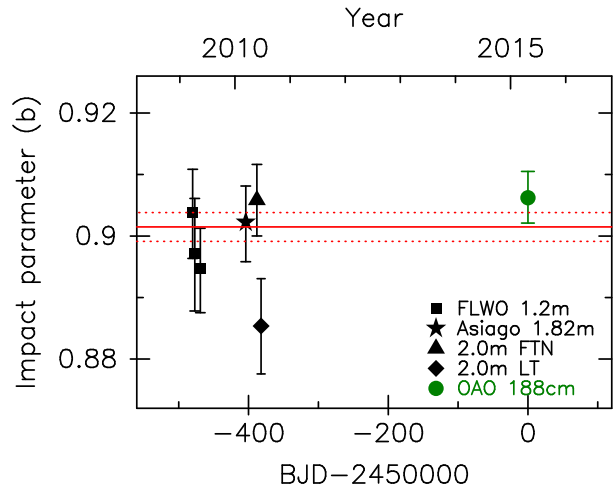}
\caption{ 
Measured impact parameter ($b$) as a function of the transit epoch. The meanings of the symbols are the same as those in Figure \ref{fig:TT}. The red solid and dotted  lines indicate the mean value and its 1-$\sigma$ boundaries, respectively ($b=0.9013\pm0.0024$).
\label{fig:b}
}
\end{center}
\end{figure}

Next we search for TDVs to check for any evidence of additional bodies.
The transit duration is generally expressed as the time interval between the first and fourth contacts of a transit, $T_{14}$, which is a function of $a/R_\mathrm{s}$, $b$, and $R_\mathrm{p}/R_\mathrm{s}$. We here assume that $a/R_\mathrm{s}$ and $R_\mathrm{p}/R_\mathrm{s}$ do not vary with time within the uncertainties and search for the time dependence of $b$ instead of $T_\mathrm{14}$. The measured $b$ values of the seven transits and their uncertainties are listed in Table \ref{tbl:TT} and are shown in Figure \ref{fig:b}. A constant fit to these values gives the mean of $b$=\rev{0.9015} $\pm$ 0.0024 with $\chi^2$=\rev{7.5} for dof=6, meaning that no significant variation is observed.

\begin{figure}
\begin{center}
\includegraphics[width=9cm]{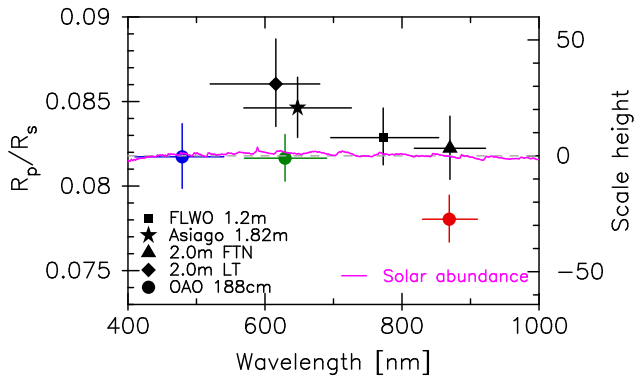}
\caption{
Measured $R_\mathrm{p}/R_\mathrm{s}$ value as a function of wavelength. The square, star, triangle, and diamond are for the data from the FLWO 1.2~m telescope/$i$-band, the Asiago 1.82~m/$R$-band, the 2.0~m FTN/$Z$-band, and the 2.0~m LT/$V$-band, respectively. The blue, green, and red circles are for the data from the OAO 188~cm telescope/MuSCAT in $g'_2$, $r'_2$, and $z_\mathrm{s,2}$ bands, respectively. The horizontal bars indicate the full width at half maximum of the response functions of the respective filters. The gray dashed and magenta solid lines represent the mean value and a solar-abundance atmospheric model, respectively.
\label{fig:RpRs}
}
\end{center}
\end{figure}

Finally, we search for $R_\mathrm{p}/R_\mathrm{s}$ variations with wavelength, which can in principle be observed due to the atmospheric opacity variations. The measured $R_\mathrm{p}/R_\mathrm{s}$ values and their uncertainties in respective bands are listed in Ta ble \ref{tbl:RpRs} and shown in Figure \ref{fig:RpRs}. A constant fit to these values gives the mean value of $R_\mathrm{p}/R_\mathrm{s}=\rev{0.08179} \pm \rev{0.00064}$ with $\chi^2$=\rev{13.0} for dof=6, meaning that $R_\mathrm{p}/R_\mathrm{s}$ is constant over the observed wavelength range within 2 $\sigma$. The consistency with a flat line can be attributed to the fact that the expected atmospheric scale height of this planet is too low. The scale height is expressed as 
\begin{eqnarray}
\label{eq:H}
H \equiv kT/\mu g_\mathrm{p},
\end{eqnarray} 
where $k$ is the Boltzmann constant, $T$ is the atmospheric temperature, $\mu$ is the mean molecular weight, and $g_\mathrm{p}$ is the planetary surface gravity. HAT-P-14b has a relatively large surface gravity of $g_\mathrm{p} = 38$~ms$^{-2}$, making $H$ too low to cause observable variations in $R_\mathrm{p}/R_\mathrm{s}$. The right vertical axis of Figure \ref{fig:RpRs} shows a scale in the unit of the expected scale height, assuming an isothermal atmosphere with the equilibrium temperature of $T=1624$~K \citep{2012MNRAS.426.1291S} and a solar abundance with $\mu$=2.35~g. The 1-$\sigma$ uncertainties of the measured $R_\mathrm{p}/R_\mathrm{s}$ is the level of $\sim$10--20 times as large as one scale height.

In addition, to compare the observed data with a theoretical atmospheric model, we calculate a model spectrum of $R_\mathrm{p}/R_\mathrm{s}$ for this planet, assuming an isothermal atmosphere with $T=1624$~K and a solar abundance. The calculation is done based on the method described in \citet{2014ApJ...790..108F}, but in this case line absorption of H$_2$S, OCS, O$_2$, TiO, and VO are taken into account in addition to those treated in \citet{2014ApJ...790..108F} (H$_2$, H$_2$O, CH$_4$, CO, CO$_2$, NH$_3$, N$_2$, Na, and K). Further, the absorption cross sections for these species are calculated on a wavelength-by-wavelength basis with a step size given by dividing the wavelength range from 0.3 to 1.0 $\mu$m by $\sim$2.5$\times10^5$ in log scale
without taking the geometric mean for a certain range of wavelengths as done in \citet{2014ApJ...790..108F}. 
An initial model spectrum is calculated by assuming that $R_0$, the planetocentric distance at which the atmospheric pressure is 10 bar, is equal to 1.219 $R_\mathrm{Jup}$ \citep{2012MNRAS.426.1291S}.
The calculated model spectrum is shown as a magenta solid line in Figure \ref{fig:RpRs} where the initial model is vertically shifted such that the mean value of the model fits to the observation. Again, the expected $R_\mathrm{p}/R_\mathrm{s}$ variations of this planet are too low with respect to the observational uncertainties.

\begin{deluxetable}{cccc}
\tablewidth{8cm}
\tablecaption{
The Measured Planet-to-star Radius Ratio ($R_\mathrm{p}/R_\mathrm{s}$) for Each Band
\label{tbl:RpRs}
}
\tablehead{
Telescope & Filter & Wavelength center & $R_\mathrm{p}/R_\mathrm{s}$\\
& & (nm) & 
}
\startdata
OAO 188cm & $g'_2$ & 479 & 0.0817 $\pm$  0.0019\\
LT & $V$ & 616 & 0.0860 $\pm$ 0.0026\\
OAO 188cm & $r'_2$ & 629 & 0.0817 $\pm$  0.0014\\
Asiago 1.82m & $R$ & 648 & 0.0846 $\pm$ 0.0018 \\
FWLO 1.2m & $i'$ & 773 & 0.0829 $\pm$ 0.0017\\
OAO 188cm & $z_\mathrm{s,2}$ & 869 & 0.0780 $\pm$ 0.0014\\
FTN & $Z$ & 870 & 0.0822 $\pm$ 0.0019
\enddata
\end{deluxetable}

\begin{figure*}
\begin{center}
\includegraphics[width=12cm]{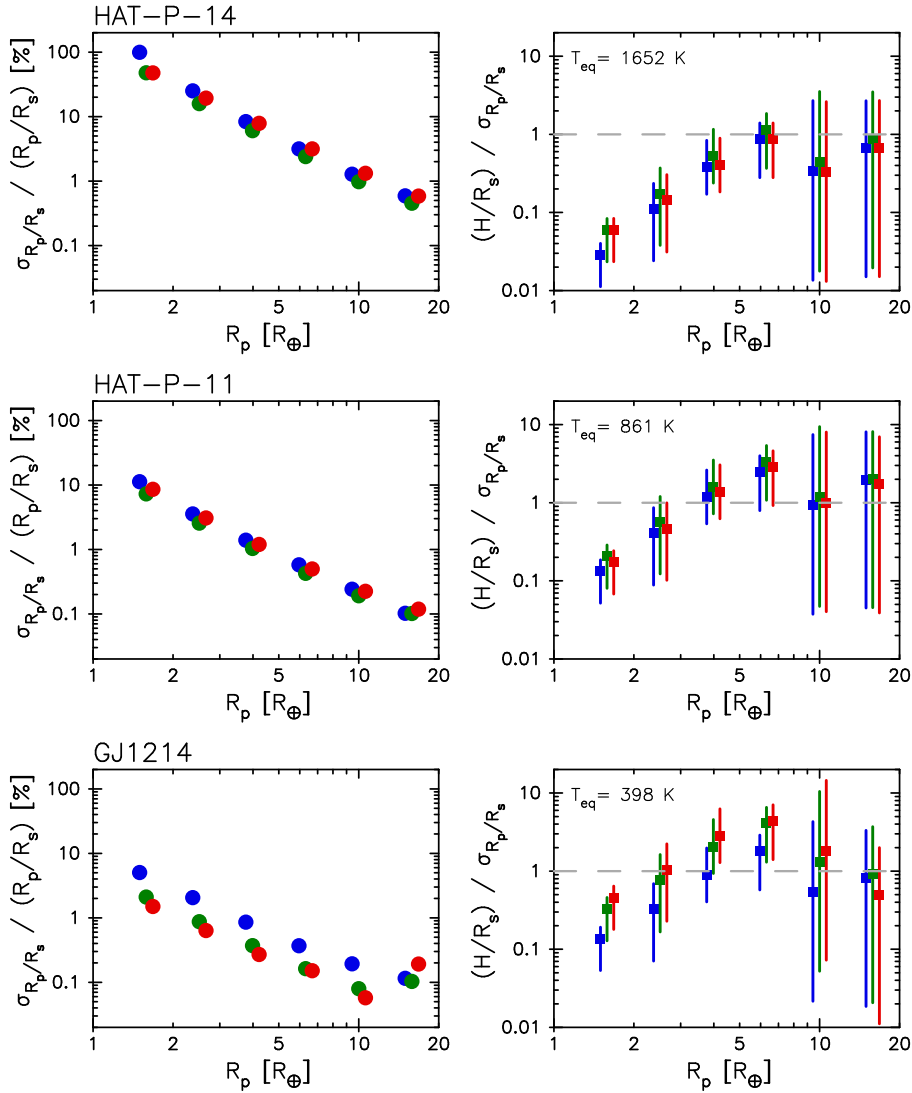}
\vspace{10pt}
\caption{
\rev{
(Left column) The fractional measurement error of $R_\mathrm{p}/R_\mathrm{s}$ that can be achieved by MuSCAT as a function of planetary radius. The top, middle, and bottom panels are for the cases that the host star is assumed to be HAT-P-14, HAT-P-11, and GJ1214, respectively. The blue, green, and red circles represents the results of the simulation for the  $g'_2$, $r'_2$, and $z_\mathrm{s,2}$ bands, respectively, where the blue and red ones are slightly shifted in the horizontal direction for clarity. In all cases a transiting planet with $P$=3 days, $b$=0.5, and $e$=0 are assumed. See Section \ref{sec:simulation} for more details.  Note that the rise at the largest $R_\mathrm{p}$ in GJ1214 is due to the planet having a grazing orbit. 
(Right column) The same as the left column, but the ratio of the expected $H/R_\mathrm{s}$ to the achievable measurement error of $R_\mathrm{p}/R_\mathrm{s}$. The value $H$ is calculated assuming a clear isothermal solar-abundance atmosphere with $\mu$=2.35~g. The adopted atmospheric temperature, assuming the bond albedo of 0.4, is indicated in each panel. The squares and vertical bars represent the median values and min-max ranges, respectively, calculated using the $g_\mathrm{p}$ distribution of the known transiting planets shown in Figure \ref{fig:gp}. When the head of a bar exceeds unity (indicated by gray dashed line) the measurement can roughly be considered to have a sensitivity to the atmosphere of the same-size planet if it has a low enough surface gravity. }
\label{fig:simulation}
}
\end{center}
\end{figure*}

\subsection{Achievable Measurement Error of $R_\mathrm{p}/R_\mathrm{s}$ with MuSCAT}
\label{sec:simulation}

The detectability of planetary atmospheric features through the transmission spectrophotometry depends on how precisely the $R_\mathrm{p}/R_\mathrm{s}$ value can be measured with respect to the expected atmospheric scale height of the planet. For planning future observations with MuSCAT, it is worthwhile investigating what types of planet can be targeted by this instrument for the atmospheric study.

To this end, we simulate the achievable measurement error of $R_\mathrm{p}/R_\mathrm{s}$ with MuSCAT for \rev{planets with a range of sizes around three types of host stars:} the $V=10$ F-dwarf HAT-P-14, the nearby (38 pc) K-dwarf HAT-P-11 \citep{2010ApJ...710.1724B}, and the nearby (14 pc) M-dwarf GJ1214 \citep{2009Natur.462..891C}. The properties of the respective stars are listed in Table \ref{tbl:mag_radius}. In each case we simulate that the host star has the same size and brightness with the assumed star. The simulation procedure is as follows. First we create a mock OOT light curve for each star and each band from the residual light curve \rev{in the same band} of HAT-P-14b (see Figure \ref{fig:corrected}). 
For mock OOT light curves of HAT-P-14 we use the residual light curves as they are. On the other hand, for HAT-P-11 and GJ1214 we assume that all the conditions except for the brightness of the host star (scintillation noise, photon noises from comparison stars, sky background noise, and systematic noises) are the same as those in the case of HAT-P-14, and set the error bar of each data point in the mock OOT light curve by modifying the error bar in the residual light curve as
\begin{eqnarray}
\sigma_\mathrm{mock} &=& \beta \sqrt{\sigma_\mathrm{ph,host}^2 + \sigma_\mathrm{others}^2}\\
&=& \beta \sqrt{ \sigma_\mathrm{ph,host}^2 + \mathrm{RMS}_\mathrm{resi}^2 - \sigma_\mathrm{ph,hatp14}^2},
\end{eqnarray}
where $\beta$ is the red noise factor given in Table \ref{tbl:data_list}, $\sigma_\mathrm{ph,host}$ is the theoretical photon noise of the host star, $\sigma_\mathrm{others}$ is the total of the noise components other than the photon noise of HAT-P-14 in the residual light curve, $\mathrm{RMS}_\mathrm{resi}$ is the rms value of the residual light curve, and $\sigma_\mathrm{ph, hatp14}$ is the theoretical photon noise of HAT-P-14.
Next we create mock transit light curves by embedding transit models in the mock OOT light curves.
We assume a typical transiting planet with $P$=3 days, $b$=0.5, and $e$=0, and range the planetary radius from $\log R_\mathrm{p}$ ($R_\oplus$)=0.2 to 1.2 with a step size of d$\log R_\mathrm{p}$=0.2.
For $w_1$ and $w_2$ we adopt the theoretical values from \citet{2011A&A...529A..75C} for each star and each band. 
Then we fit the mock transit light curves with a transit model to estimate the 1$\sigma$ uncertainty of the measured $R_\mathrm{p}/R_\mathrm{s}$, using the AMOEBA algorithm with the same parameterization as in Section \ref{sec:fit_muscat}. At this time we fit the $g'_2$-, $r'_2$-, and $z_\mathrm{s,2}$-band light curves simultaneously by treating $R_\mathrm{p}/R_\mathrm{s}$, \rev{$w_1$, $w_2$}, $k_0$, and $k_t$ as independent free parameters for each band and  $T_\mathrm{c}$ as a common free parameter for all bands. We fix $b$ and $a/R_\mathrm{s}$ at the input values, assuming that these parameters are well determined beforehand somewhere. We also impose priors on $w_1$ and $w_2$ in the same way as in Section \ref{sec:fit_muscat} and \ref{sec:fit_all}.
In the left column of Figure \ref{fig:simulation}, we plot the derived fractional 1-$\sigma$ uncertainty of $R_\mathrm{p}/R_\mathrm{s}$ as a function of planetary radius. The blue, green, and red circles represent the results for the $g'_2$, $r'_2$, and $z_\mathrm{s,2}$ bands, respectively. 
In the cases of HAT-P-14 and HAT-P-11, the fractional $R_\mathrm{p}/R_\mathrm{s}$ uncertainties in the three bands are all similar because the photometric errors are limited by the wavelength-independent scintillation noise. On the other hand, in the case of GJ1214, the fractional $R_\mathrm{p}/R_\mathrm{s}$ uncertainty in $g'_2$ band is worse than the other two because the photometric errors are limited by the photon noise of the red host star. We note that a small amount of changes of the input values of $P$ and $b$, for example,  in the ranges of $2 < P (\mathrm{days}) < 5$ and $0.4 < b < 0.6$, do not change the results so much and provide the same conclusion described later.

\begin{figure}
\begin{center}
\includegraphics[width=8cm]{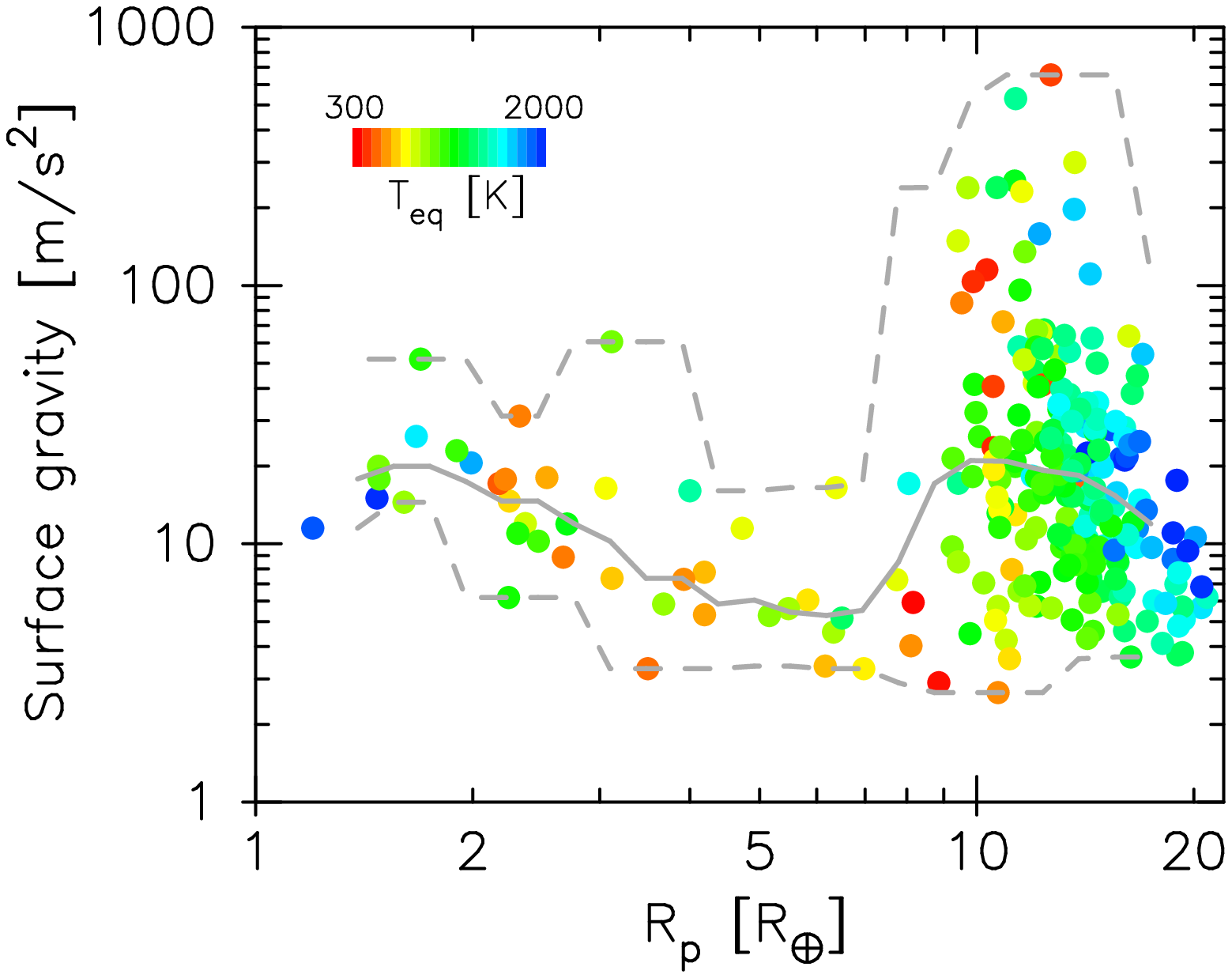}
\caption{
\rev{Planetary surface gravity as a function of planetary radius of the known transiting planets with the mass measured with the accuracy better than 30\% and with the equilibrium temperature ranging from $T_\mathrm{eq}=300$ to 2000~K. The estimated $T_\mathrm{eq}$ of each planet, assuming the planetary bond albedo of 0.4, is indicated by color. The data are taken from http://exoplanet.eu, with some erroneous data eliminated. The gray solid and lower (upper) dashed lines indicate the moving median and minimum (maximum) values with the window size of d$\log (R_\mathrm{p})=0.2$, which are used for estimating the range of expected scale height in Figure \ref{fig:simulation}.}
\label{fig:gp}
}
\end{center}
\end{figure}

\begin{deluxetable}{cccc}
\tablewidth{8.5cm}
\tablecaption{Properties of the Host Stars Assumed for the Simulation of Achievable Measurement Error of $R_\mathrm{p}/R_\mathrm{s}$
\label{tbl:mag_radius}}
\tablehead{
& \colhead{HAT-P-14} & \colhead{HAT-P-11}& \colhead{GJ1214}
}
\startdata
Spec. Type & F5V  & K4V & M4.5V \\
Radius (R$_\odot$) &  1.59 & 0.683 & 0.211 \\
$T_\mathrm{eff}$ (K) & 6600 & 4780 & 3026 \\
Distance (pc) & 205 & 38 & 14.3 \\
$g'$ & 10.18 & 9.97\tablenotemark{a}& 15.58\tablenotemark{b}\\
$r'$ & 9.94  & 9.16\tablenotemark{a}& 14.08\tablenotemark{b}\\
$z'$ &  10.30 & 8.79\tablenotemark{a}& 11.86\tablenotemark{b}\\
References & (1), (2), (3), (4) & (5), (6) & (7), (8)
\enddata
\tablecomments{
\tablenotetext{a}{\ Converted from $T_\mathrm{eff}$, $J$=7.608 \citep[2MASS,][]{2003yCat.2246....0C}, and $V$=9.47 \citep[Tycho-2,][]{2000A&A...355L..27H} via the color-$T_\mathrm{eff}$ relations of \citet{2013ApJ...771...40B}.}
\tablenotetext{b}{\ Converted from $T_\mathrm{eff}$, $J$=9.750 (2MASS) and $V$=14.71 \citep{2011ApJ...736...12B} via the color-$T_\mathrm{eff}$ relations of \citet{2013ApJ...771...40B}.}
}
\tablerefs{
(1) \cite{2011AJ....141..161S}, (2) \citet{2012MNRAS.426.1291S}, (3) \citet{2010ApJ...715..458T}, (4) the SDSS catalog \citep{2012ApJS..203...21A}, (5) \citet{2010ApJ...710.1724B}, (6) \citet{2011ApJ...740...33D}, (7) \citet{2014AJ....148...91L}, (8) \citet{2009Natur.462..891C}.}
\end{deluxetable}

Next we estimate how sensitive these $R_\mathrm{p}/R_\mathrm{s}$ measurements are to atmospheric signatures by comparing the estimated 1-$\sigma$ measurement error of $R_\mathrm{p}/R_\mathrm{s}$,  $\sigma_{R_\mathrm{p}/R_\mathrm{s}}$, with the expected variation of $R_\mathrm{p}/R_\mathrm{s}$ due to the atmospheric effect. When a planet has an atmosphere with no thick clouds, $R_\mathrm{p}/R_\mathrm{s}$ can vary with wavelength by up to several $\times  H/R_\mathrm{s}$. Therefore, if $\sigma_{R_\mathrm{p}/R_\mathrm{s}}$ is comparable to or smaller than the expected value of $H/R_\mathrm{s}$, then it can roughly be considered that the measurement is sensitive to the atmosphere. To estimate $H$, which is given by Equation (\ref{eq:H}), we assume a clear isothermal solar-abundance ($\mu$=2.35 g) atmosphere with the temperature equal to the equilibrium temperature, $T_\mathrm{eq}$, of a planet with $P$=3 days and bond albedo of $A$=0.4: $T_\mathrm{eq}$ for a planet around HAT-P-14, HAT-P-11, and GJ1214 is calculated to be 1652 K, 861 K, and 398 K, respectively. 
We note that $T_\mathrm{eq}$ can vary by up to $\sim$15~\% in the range of  2 $< P \mathrm{(days)} < 5$  and $0 < A < 0.7$; however, the impact of this change on $H$ is trivial compared with that of the possible range of the surface gravity $g_\mathrm{p}$, which extends by a factor of several or even two orders of magnitude depending on the planetary size (see Figure \ref{fig:gp}). Therefore, we do not consider other possibilities for $P$ and $A$.
For the possible values of $g_\mathrm{p}$, we use the observed distribution from the known transiting planets shown in Figure \ref{fig:gp}, with the mass measured with the accuracy better than 30\%, and with $T_\mathrm{eq}$ ranging from 300 to 2000~K.
Then, using these values we calculate the ratio of $H/R_\mathrm{s}$ to $\sigma_{R_\mathrm{p}/R_\mathrm{s}}$ as a function of planetary radius as shown in the right column of Figure \ref{fig:simulation}. In this plot the squares and vertical bars represent the values and ranges that are calculated from the median values and min-max ranges, respectively, of the $g_\mathrm{p}$ distribution within the range of $\log R_\mathrm{p} \pm 0.1$.  
\rev{If a part of the vertical var exceeds unity (indicated by gray dashed line), then it implies that the measurement has a sensitivity to the atmosphere of at least the lowest-surface gravity planets.}
In this sense, we find that MuSCAT is sensitive to the atmospheres of  planets as small as a sub-Jupiter ($R_\mathrm{p} \gtrsim 6 R_\oplus$) around HAT-P-14 in all bands, a Neptune ($\sim 4 R_\oplus$) around HAT-P-11 in all bands, and a super-Earth ($\sim 2.5 R_\oplus$) around GJ1214 in $r'_2$ and $z_\mathrm{s,2}$ bands.
Because the limiting factor for the case of GJ1214 is the photon noise of the host star, 
even the atmospheres of planets smaller than $\sim2.5 R_\oplus$ could be probed if such planets will be discovered around  much closer (brighter) mid-to-late M dwarfs.
 
This sensitivity of MuSCAT meets the demands in the K2 and TESS era.
The K2 and TESS missions are expected to discover hundreds of transiting super-Earths and Neptunes around nearby M dwarfs in the coming years \citep{2014PASP..126..398H,2015ApJ...809...77S}, which are potentially attractive targets for the atmospheric study. However, a certain fraction of these planets could be covered by thick clouds which will prevent us from detecting molecular features \citep[like the case of GJ1214b,][]{2014Natur.505...69K}. MuSCAT can thus play an important role in the search for cloud-free planets from a number of the to-be-discovered nearby super-Earths and Neptunes. 
Although the $g'_2$ band does not have a sensitivity to the atmosphere of a super-Earth even around a GJ1214-like star, it is still useful to validate planetary candidates. This is because in most cases the wavelength-dependent transit-depth variations due to a contamination of eclipsing binary will be much larger than those of the atmospheric origin. In particular, the $g'_2$ band covers a completely different wavelength range from (\rev{covers} the lower edge of) the spectral response of TESS ({\it Kepler}), making it valuable for the planet validation.

\section{Summary}
\label{sec:summary}

We present the results of a pilot observation with MuSCAT, a multiband imager newly developed for the OAO 188~cm telescope aimed at studying transiting exoplanets. We observed a primary transit of HAT-P-14b, a hot Jupiter orbiting a $V=10$ star, through the SDSS $g'_2$-, $r'_2$-, and $z_\mathrm{s,2}$-band filters. From the 2.9~hr observation, we have achieved the five minutes binned photometric precisions of 0.028\%, 0.022\%, and 0.024\% in the $g'_2$, $r'_2$, and $z_\mathrm{s,2}$ bands, respectively, providing the best-quality transit data for this planet. 

From a homogeneous analysis of the data including six published light curves, we search for TTVs, TDVs, and $R_\mathrm{p}/R_\mathrm{s}$ variations with wavelength, but can find no noticeable variation in any parameters. On the other hand, using the transit-subtracted light curves, we estimate achievable measurement error of $R_\mathrm{p}/R_\mathrm{s}$ for various planetary sizes assuming three types of host stars, namely HAT-P-14, HAT-P-11, and GJ1214.
Comparing our results with the expected atmospheric scale heights of planets with the lowest surface gravity, we find that MuSCAT is capable of probing the atmospheres of planets as small as a sub-Jupiter around HAT-P-14 in all bands, a Neptune around HAT-P-11 in all bands, and a super-Earth around GJ1214 in $r'_2$ and $z_\mathrm{s}$ bands. 
These results promise that MuSCAT will bear a lot of fruits in the K2 and TESS era.

\acknowledgements
We thank Elaine Simpson for kindly providing the FTN and LT light curves. We also thank Iain Steele for the filter information of LT/RISE.
A.F. acknowledges supports by the Astrobiology Center Project of  National Institutes of Natural Sciences (NINS) (Grant Number AB271009).
N.N. acknowledges supports by the NAOJ Fellowship, Inoue Science Research Award, and grant-in-aid for Scientific Research (A) (JSPS KAKENHI Grant Number 25247026).
Y.K. acknowledges supports by grant-in-aid for JSPS Fellows (JSPS KAKENHI Grant Number 15J08463) and the Leading Graduate Course for Frontiers of Mathematical Sciences and Physics.

\bibliographystyle{apj}
\bibliography{ref}

\begin{thebibliography}{}
\expandafter\ifx\csname natexlab\endcsname\relax\def\natexlab#1{#1}\fi

\bibitem[{{Ahn} {et~al.}(2012){Ahn}, {Alexandroff}, {Allende Prieto},
  {Anderson}, {Anderton}, {Andrews}, {Aubourg}, {Bailey}, {Balbinot}, {Barnes},
  \& et~al.}]{2012ApJS..203...21A}
{Ahn}, C.~P., {Alexandroff}, R., {Allende Prieto}, C., {et~al.} 2012, \apjs,
  203, 21

\bibitem[{{Bakos} {et~al.}(2010){Bakos}, {Torres}, {P{\'a}l}, {Hartman},
  {Kov{\'a}cs}, {Noyes}, {Latham}, {Sasselov}, {Sip{\H o}cz}, {Esquerdo},
  {Fischer}, {Johnson}, {Marcy}, {Butler}, {Isaacson}, {Howard}, {Vogt},
  {Kov{\'a}cs}, {Fernandez}, {Mo{\'o}r}, {Stefanik}, {L{\'a}z{\'a}r}, {Papp},
  \& {S{\'a}ri}}]{2010ApJ...710.1724B}
{Bakos}, G.~{\'A}., {Torres}, G., {P{\'a}l}, A., {et~al.} 2010, \apj, 710, 1724

\bibitem[{{Barros} {et~al.}(2015){Barros}, {Almenara}, {Demangeon}, {Tsantaki},
  {Santerne}, {Armstrong}, {Barrado}, {Brown}, {Deleuil}, {Lillo-Box},
  {Osborn}, {Pollacco}, {Abe}, {Andre}, {Bendjoya}, {Boisse}, {Bonomo},
  {Bouchy}, {Bruno}, {Cerda}, {Courcol}, {D{\'{\i}}az}, {H{\'e}brard}, {Kirk},
  {Lachuri{\'e}}, {Lam}, {Martinez}, {McCormac}, {Moutou}, {Rajpurohit},
  {Rivet}, {Spake}, {Suarez}, {Toublanc}, \& {Walker}}]{2015MNRAS.454.4267B}
{Barros}, S.~C.~C., {Almenara}, J.~M., {Demangeon}, O., {et~al.} 2015, \mnras,
  454, 4267

\bibitem[{{Berta} {et~al.}(2011){Berta}, {Charbonneau}, {Bean}, {Irwin},
  {Burke}, {D{\'e}sert}, {Nutzman}, \& {Falco}}]{2011ApJ...736...12B}
{Berta}, Z.~K., {Charbonneau}, D., {Bean}, J., {et~al.} 2011, \apj, 736, 12

\bibitem[{{Boyajian} {et~al.}(2013){Boyajian}, {von Braun}, {van Belle},
  {Farrington}, {Schaefer}, {Jones}, {White}, {McAlister}, {ten Brummelaar},
  {Ridgway}, {Gies}, {Sturmann}, {Sturmann}, {Turner}, {Goldfinger}, \&
  {Vargas}}]{2013ApJ...771...40B}
{Boyajian}, T.~S., {von Braun}, K., {van Belle}, G., {et~al.} 2013, \apj, 771,
  40

\bibitem[{{Charbonneau} {et~al.}(2009){Charbonneau}, {Berta}, {Irwin}, {Burke},
  {Nutzman}, {Buchhave}, {Lovis}, {Bonfils}, {Latham}, {Udry}, {Murray-Clay},
  {Holman}, {Falco}, {Winn}, {Queloz}, {Pepe}, {Mayor}, {Delfosse}, \&
  {Forveille}}]{2009Natur.462..891C}
{Charbonneau}, D., {Berta}, Z.~K., {Irwin}, J., {et~al.} 2009, \nat, 462, 891

\bibitem[{{Claret}(2009)}]{2009A&A...506.1335C}
{Claret}, A. 2009, \aap, 506, 1335

\bibitem[{{Claret} \& {Bloemen}(2011)}]{2011A&A...529A..75C}
{Claret}, A., \& {Bloemen}, S. 2011, \aap, 529, A75

\bibitem[{{Claret} {et~al.}(2013){Claret}, {Hauschildt}, \&
  {Witte}}]{2013A&A...552A..16C}
{Claret}, A., {Hauschildt}, P.~H., \& {Witte}, S. 2013, \aap, 552, A16

\bibitem[{{Col{\'o}n} \& {Ford}(2011)}]{2011PASP..123.1391C}
{Col{\'o}n}, K.~D., \& {Ford}, E.~B. 2011, \pasp, 123, 1391

\bibitem[{{Croll} {et~al.}(2011){Croll}, {Albert}, {Jayawardhana},
  {Miller-Ricci Kempton}, {Fortney}, {Murray}, \&
  {Neilson}}]{2011ApJ...736...78C}
{Croll}, B., {Albert}, L., {Jayawardhana}, R., {et~al.} 2011, \apj, 736, 78

\bibitem[{{Cutri} {et~al.}(2003){Cutri}, {Skrutskie}, {van Dyk}, {Beichman},
  {Carpenter}, {Chester}, {Cambresy}, {Evans}, {Fowler}, {Gizis}, {Howard},
  {Huchra}, {Jarrett}, {Kopan}, {Kirkpatrick}, {Light}, {Marsh}, {McCallon},
  {Schneider}, {Stiening}, {Sykes}, {Weinberg}, {Wheaton}, {Wheelock}, \&
  {Zacarias}}]{2003yCat.2246....0C}
{Cutri}, R.~M., {Skrutskie}, M.~F., {van Dyk}, S., {et~al.} 2003, VizieR Online
  Data Catalog, 2246, 0

\bibitem[{{de Mooij} {et~al.}(2012){de Mooij}, {Brogi}, {de Kok},
  {Koppenhoefer}, {Nefs}, {Snellen}, {Greiner}, {Hanse}, {Heinsbroek}, {Lee},
  \& {van der Werf}}]{2012A&A...538A..46D}
{de Mooij}, E.~J.~W., {Brogi}, M., {de Kok}, R.~J., {et~al.} 2012, \aap, 538,
  A46

\bibitem[{{de Mooij} {et~al.}(2013){de Mooij}, {Brogi}, {de Kok}, {Snellen},
  {Croll}, {Jayawardhana}, {Hoekstra}, {Otten}, {Bekkers}, {Haffert}, \& {van
  Houdt}}]{2013ApJ...771..109D}
---. 2013, \apj, 771, 109

\bibitem[{{Deming} {et~al.}(2011){Deming}, {Sada}, {Jackson}, {Peterson},
  {Agol}, {Knutson}, {Jennings}, {Haase}, \& {Bays}}]{2011ApJ...740...33D}
{Deming}, D., {Sada}, P.~V., {Jackson}, B., {et~al.} 2011, \apj, 740, 33

\bibitem[{{Dravins} {et~al.}(1998){Dravins}, {Lindegren}, {Mezey}, \&
  {Young}}]{1998PASP..110..610D}
{Dravins}, D., {Lindegren}, L., {Mezey}, E., \& {Young}, A.~T. 1998, \pasp,
  110, 610

\bibitem[{{Eastman} {et~al.}(2010){Eastman}, {Siverd}, \&
  {Gaudi}}]{2010PASP..122..935E}
{Eastman}, J., {Siverd}, R., \& {Gaudi}, B.~S. 2010, \pasp, 122, 935

\bibitem[{{Fukui} {et~al.}(2011){Fukui}, {Narita}, {Tristram}, {Sumi}, {Abe},
  {Itow}, {Sullivan}, {Bond}, {Hirano}, {Tamura}, {Bennett}, {Furusawa},
  {Hayashi}, {Hearnshaw}, {Hosaka}, {Kamiya}, {Kobara}, {Korpela}, {Kilmartin},
  {Lin}, {Ling}, {Makita}, {Masuda}, {Matsubara}, {Miyake}, {Muraki}, {Nagaya},
  {Nishimoto}, {Ohnishi}, {Omori}, {Perrott}, {Rattenbury}, {Saito}, {Skuljan},
  {Suzuki}, {Sweatman}, \& {Wada}}]{2011PASJ...63..287F}
{Fukui}, A., {Narita}, N., {Tristram}, P.~J., {et~al.} 2011, \pasj, 63, 287

\bibitem[{{Fukui} {et~al.}(2013){Fukui}, {Narita}, {Kurosaki}, {Ikoma},
  {Yanagisawa}, {Kuroda}, {Shimizu}, {Takahashi}, {Ohnuki}, {Onitsuka},
  {Hirano}, {Suenaga}, {Kawauchi}, {Nagayama}, {Ohta}, {Yoshida}, {Kawai}, \&
  {Izumiura}}]{2013ApJ...770...95F}
{Fukui}, A., {Narita}, N., {Kurosaki}, K., {et~al.} 2013, \apj, 770, 95

\bibitem[{{Fukui} {et~al.}(2014){Fukui}, {Kawashima}, {Ikoma}, {Narita},
  {Onitsuka}, {Ita}, {Onozato}, {Nishiyama}, {Baba}, {Ryu}, {Hirano}, {Hori},
  {Kurosaki}, {Kawauchi}, {Takahashi}, {Nagayama}, {Tamura}, {Kawai}, {Kuroda},
  {Nagayama}, {Ohta}, {Shimizu}, {Yanagisawa}, {Yoshida}, \&
  {Izumiura}}]{2014ApJ...790..108F}
{Fukui}, A., {Kawashima}, Y., {Ikoma}, M., {et~al.} 2014, \apj, 790, 108

\bibitem[{{Gibson}(2014)}]{2014MNRAS.445.3401G}
{Gibson}, N.~P. 2014, \mnras, 445, 3401

\bibitem[{{Gibson} {et~al.}(2012){Gibson}, {Aigrain}, {Roberts}, {Evans},
  {Osborne}, \& {Pont}}]{2012MNRAS.419.2683G}
{Gibson}, N.~P., {Aigrain}, S., {Roberts}, S., {et~al.} 2012, \mnras, 419, 2683

\bibitem[{{H{\o}g} {et~al.}(2000){H{\o}g}, {Fabricius}, {Makarov}, {Urban},
  {Corbin}, {Wycoff}, {Bastian}, {Schwekendiek}, \&
  {Wicenec}}]{2000A&A...355L..27H}
{H{\o}g}, E., {Fabricius}, C., {Makarov}, V.~V., {et~al.} 2000, \aap, 355, L27

\bibitem[{{Howell} {et~al.}(2014){Howell}, {Sobeck}, {Haas}, {Still},
  {Barclay}, {Mullally}, {Troeltzsch}, {Aigrain}, {Bryson}, {Caldwell},
  {Chaplin}, {Cochran}, {Huber}, {Marcy}, {Miglio}, {Najita}, {Smith},
  {Twicken}, \& {Fortney}}]{2014PASP..126..398H}
{Howell}, S.~B., {Sobeck}, C., {Haas}, M., {et~al.} 2014, \pasp, 126, 398

\bibitem[{{Knutson} {et~al.}(2014){Knutson}, {Fulton}, {Montet}, {Kao}, {Ngo},
  {Howard}, {Crepp}, {Hinkley}, {Bakos}, {Batygin}, {Johnson}, {Morton}, \&
  {Muirhead}}]{2014ApJ...785..126K}
{Knutson}, H.~A., {Fulton}, B.~J., {Montet}, B.~T., {et~al.} 2014, \apj, 785,
  126

\bibitem[{{Kreidberg} {et~al.}(2014){Kreidberg}, {Bean}, {D{\'e}sert},
  {Benneke}, {Deming}, {Stevenson}, {Seager}, {Berta-Thompson}, {Seifahrt}, \&
  {Homeier}}]{2014Natur.505...69K}
{Kreidberg}, L., {Bean}, J.~L., {D{\'e}sert}, J.-M., {et~al.} 2014, \nat, 505,
  69

\bibitem[{{Lurie} {et~al.}(2014){Lurie}, {Henry}, {Jao}, {Quinn}, {Winters},
  {Ianna}, {Koerner}, {Riedel}, \& {Subasavage}}]{2014AJ....148...91L}
{Lurie}, J.~C., {Henry}, T.~J., {Jao}, W.-C., {et~al.} 2014, \aj, 148, 91

\bibitem[{{Mancini} {et~al.}(2013){Mancini}, {Ciceri}, {Chen}, {Tregloan-Reed},
  {Fortney}, {Southworth}, {Tan}, {Burgdorf}, {Calchi Novati}, {Dominik},
  {Fang}, {Finet}, {Gerner}, {Hardis}, {Hinse}, {J{\o}rgensen}, {Liebig},
  {Nikolov}, {Ricci}, {Sch{\"a}fer}, {Sch{\"o}nebeck}, {Skottfelt}, {Wertz},
  {Alsubai}, {Bozza}, {Browne}, {Dodds}, {Gu}, {Harps{\o}e}, {Henning},
  {Hundertmark}, {Jessen-Hansen}, {Kains}, {Kerins}, {Kjeldsen}, {Lund},
  {Lundkvist}, {Madhusudhan}, {Mathiasen}, {Penny}, {Prof}, {Rahvar}, {Sahu},
  {Scarpetta}, {Snodgrass}, \& {Surdej}}]{2013MNRAS.436....2M}
{Mancini}, L., {Ciceri}, S., {Chen}, G., {et~al.} 2013, \mnras, 436, 2

\bibitem[{{Mancini} {et~al.}(2014){Mancini}, {Southworth}, {Ciceri}, {Dominik},
  {Henning}, {J{\o}rgensen}, {Lanza}, {Rabus}, {Snodgrass}, {Vilela},
  {Alsubai}, {Bozza}, {Bramich}, {Calchi Novati}, {D'Ago}, {Figuera Jaimes},
  {Galianni}, {Gu}, {Harps{\o}e}, {Hinse}, {Hundertmark}, {Juncher}, {Kains},
  {Korhonen}, {Popovas}, {Rahvar}, {Skottfelt}, {Street}, {Surdej}, {Tsapras},
  {Wang}, \& {Wertz}}]{2014A&A...562A.126M}
{Mancini}, L., {Southworth}, J., {Ciceri}, S., {et~al.} 2014, \aap, 562, A126

\bibitem[{{Mandel} \& {Agol}(2002)}]{2002ApJ...580L.171M}
{Mandel}, K., \& {Agol}, E. 2002, \apjl, 580, L171

\bibitem[{{Mislis} {et~al.}(2010){Mislis}, {Schr{\"o}ter}, {Schmitt}, {Cordes},
  \& {Reif}}]{2010A&A...510A.107M}
{Mislis}, D., {Schr{\"o}ter}, S., {Schmitt}, J.~H.~M.~M., {Cordes}, O., \&
  {Reif}, K. 2010, \aap, 510, A107

\bibitem[{{Narita} {et~al.}(2013{\natexlab{a}}){Narita}, {Nagayama}, {Suenaga},
  {Fukui}, {Ikoma}, {Nakajima}, {Nishiyama}, \& {Tamura}}]{2013PASJ...65...27N}
{Narita}, N., {Nagayama}, T., {Suenaga}, T., {et~al.} 2013{\natexlab{a}},
  \pasj, 65, 27

\bibitem[{{Narita} {et~al.}(2007){Narita}, {Enya}, {Sato}, {Ohta}, {Winn},
  {Suto}, {Taruya}, {Turner}, {Aoki}, {Yoshii}, {Yamada}, \&
  {Tamura}}]{2007PASJ...59..763N}
{Narita}, N., {Enya}, K., {Sato}, B., {et~al.} 2007, \pasj, 59, 763

\bibitem[{{Narita} {et~al.}(2013{\natexlab{b}}){Narita}, {Fukui}, {Ikoma},
  {Hori}, {Kurosaki}, {Kawashima}, {Nagayama}, {Onitsuka}, {Sukom}, {Nakajima},
  {Tamura}, {Kuroda}, {Yanagisawa}, {Hirano}, {Kawauchi}, {Kuzuhara}, {Ohnuki},
  {Suenaga}, {Takahashi}, {Izumiura}, {Kawai}, \&
  {Yoshida}}]{2013ApJ...773..144N}
{Narita}, N., {Fukui}, A., {Ikoma}, M., {et~al.} 2013{\natexlab{b}}, \apj, 773,
  144

\bibitem[{{Narita} {et~al.}(2015{\natexlab{a}}){Narita}, {Hirano}, {Fukui},
  {Hori}, {Sanchis-Ojeda}, {Winn}, {Ryu}, {Kusakabe}, {Kudo}, {Onitsuka},
  {Delrez}, {Gillon}, {Jehin}, {McCormac}, {Holman}, {Izumiura}, {Takeda},
  {Tamura}, \& {Yanagisawa}}]{2015ApJ...815...47N}
{Narita}, N., {Hirano}, T., {Fukui}, A., {et~al.} 2015{\natexlab{a}}, \apj,
  815, 47

\bibitem[{{Narita} {et~al.}(2015{\natexlab{b}}){Narita}, {Fukui}, {Kusakabe},
  {Onitsuka}, {Ryu}, {Yanagisawa}, {Izumiura}, {Tamura}, \&
  {Yamamuro}}]{2015JATIS...1d5001N}
{Narita}, N., {Fukui}, A., {Kusakabe}, N., {et~al.} 2015{\natexlab{b}}, Journal
  of Astronomical Telescopes, Instruments, and Systems, 1, 045001

\bibitem[{{Nascimbeni} {et~al.}(2011){Nascimbeni}, {Piotto}, {Bedin}, \&
  {Damasso}}]{2011AA...527A..85N}
{Nascimbeni}, V., {Piotto}, G., {Bedin}, L.~R., \& {Damasso}, M. 2011, \aap,
  527, A85

\bibitem[{{Nascimbeni} {et~al.}(2013){Nascimbeni}, {Piotto}, {Pagano},
  {Scandariato}, {Sani}, \& {Fumana}}]{2013A&A...559A..32N}
{Nascimbeni}, V., {Piotto}, G., {Pagano}, I., {et~al.} 2013, \aap, 559, A32

\bibitem[{{Nascimbeni} {et~al.}(2015){Nascimbeni}, {Mallonn}, {Scandariato},
  {Pagano}, {Piotto}, {Micela}, {Messina}, {Leto}, {Strassmeier}, {Bisogni}, \&
  {Speziali}}]{2015A&A...579A.113N}
{Nascimbeni}, V., {Mallonn}, M., {Scandariato}, G., {et~al.} 2015, \aap, 579,
  A113

\bibitem[{{O'Donovan} {et~al.}(2006){O'Donovan}, {Charbonneau}, {Torres},
  {Mandushev}, {Dunham}, {Latham}, {Alonso}, {Brown}, {Esquerdo}, {Everett}, \&
  {Creevey}}]{2006ApJ...644.1237O}
{O'Donovan}, F.~T., {Charbonneau}, D., {Torres}, G., {et~al.} 2006, \apj, 644,
  1237

\bibitem[{{Ohta} {et~al.}(2009){Ohta}, {Taruya}, \&
  {Suto}}]{2009ApJ...690....1O}
{Ohta}, Y., {Taruya}, A., \& {Suto}, Y. 2009, \apj, 690, 1

\bibitem[{{P{\'a}l}(2008)}]{2008MNRAS.390..281P}
{P{\'a}l}, A. 2008, \mnras, 390, 281

\bibitem[{{Pont} {et~al.}(2008){Pont}, {Knutson}, {Gilliland}, {Moutou}, \&
  {Charbonneau}}]{2008MNRAS.385..109P}
{Pont}, F., {Knutson}, H., {Gilliland}, R.~L., {Moutou}, C., \& {Charbonneau},
  D. 2008, \mnras, 385, 109

\bibitem[{{Pont} {et~al.}(2006){Pont}, {Zucker}, \&
  {Queloz}}]{2006MNRAS.373..231P}
{Pont}, F., {Zucker}, S., \& {Queloz}, D. 2006, \mnras, 373, 231

\bibitem[{{Press} {et~al.}(1992){Press}, {Teukolsky}, {Vetterling}, \&
  {Flannery}}]{1992nrca.book.....P}
{Press}, W.~H., {Teukolsky}, S.~A., {Vetterling}, W.~T., \& {Flannery}, B.~P.
  1992, {Numerical recipes in C. The art of scientific computing}, ed. {Press,
  W.~H., Teukolsky, S.~A., Vetterling, W.~T., \& Flannery, B.~P. }

\bibitem[{{Ricker} {et~al.}(2015){Ricker}, {Winn}, {Vanderspek}, {Latham},
  {Bakos}, {Bean}, {Berta-Thompson}, {Brown}, {Buchhave}, {Butler}, {Butler},
  {Chaplin}, {Charbonneau}, {Christensen-Dalsgaard}, {Clampin}, {Deming},
  {Doty}, {De Lee}, {Dressing}, {Dunham}, {Endl}, {Fressin}, {Ge}, {Henning},
  {Holman}, {Howard}, {Ida}, {Jenkins}, {Jernigan}, {Johnson}, {Kaltenegger},
  {Kawai}, {Kjeldsen}, {Laughlin}, {Levine}, {Lin}, {Lissauer}, {MacQueen},
  {Marcy}, {McCullough}, {Morton}, {Narita}, {Paegert}, {Palle}, {Pepe},
  {Pepper}, {Quirrenbach}, {Rinehart}, {Sasselov}, {Sato}, {Seager},
  {Sozzetti}, {Stassun}, {Sullivan}, {Szentgyorgyi}, {Torres}, {Udry}, \&
  {Villasenor}}]{2015JATIS...1a4003R}
{Ricker}, G.~R., {Winn}, J.~N., {Vanderspek}, R., {et~al.} 2015, Journal of
  Astronomical Telescopes, Instruments, and Systems, 1, 014003

\bibitem[{{Schwarz}(1978)}]{1978Schwarz}
{Schwarz}, G. 1978, Ann. Statistics, 6, 461

\bibitem[{{Simpson} {et~al.}(2011){Simpson}, {Barros}, {Brown}, {Collier
  Cameron}, {Pollacco}, {Skillen}, {Stempels}, {Boisse}, {Faedi},
  {H{\'e}brard}, {McCormac}, {Sorensen}, {Street}, {Anderson}, {Bento},
  {Bouchy}, {Butters}, {Enoch}, {Haswell}, {Hebb}, {Hellier}, {Holmes},
  {Horne}, {Keenan}, {Lister}, {Maxted}, {Miller}, {Moulds}, {Moutou},
  {Norton}, {Parley}, {Santerne}, {Smalley}, {Smith}, {Todd}, {Watson}, {West},
  \& {Wheatley}}]{2011AJ....141..161S}
{Simpson}, E.~K., {Barros}, S.~C.~C., {Brown}, D.~J.~A., {et~al.} 2011, \aj,
  141, 161

\bibitem[{{Siverd} {et~al.}(2012){Siverd}, {Beatty}, {Pepper}, {Eastman},
  {Collins}, {Bieryla}, {Latham}, {Buchhave}, {Jensen}, {Crepp}, {Street},
  {Stassun}, {Gaudi}, {Berlind}, {Calkins}, {DePoy}, {Esquerdo}, {Fulton},
  {F{\H u}r{\'e}sz}, {Geary}, {Gould}, {Hebb}, {Kielkopf}, {Marshall}, {Pogge},
  {Stanek}, {Stefanik}, {Szentgyorgyi}, {Trueblood}, {Trueblood}, {Stutz}, \&
  {van Saders}}]{2012ApJ...761..123S}
{Siverd}, R.~J., {Beatty}, T.~G., {Pepper}, J., {et~al.} 2012, \apj, 761, 123

\bibitem[{{Snellen} {et~al.}(2009){Snellen}, {Koppenhoefer}, {van der Burg},
  {Dreizler}, {Greiner}, {de Hoon}, {Husser}, {Kr{\"u}hler}, {Saglia}, \&
  {Vuijsje}}]{2009A&A...497..545S}
{Snellen}, I.~A.~G., {Koppenhoefer}, J., {van der Burg}, R.~F.~J., {et~al.}
  2009, \aap, 497, 545

\bibitem[{{Southworth}(2012)}]{2012MNRAS.426.1291S}
{Southworth}, J. 2012, \mnras, 426, 1291

\bibitem[{{Sullivan} {et~al.}(2015){Sullivan}, {Winn}, {Berta-Thompson},
  {Charbonneau}, {Deming}, {Dressing}, {Latham}, {Levine}, {McCullough},
  {Morton}, {Ricker}, {Vanderspek}, \& {Woods}}]{2015ApJ...809...77S}
{Sullivan}, P.~W., {Winn}, J.~N., {Berta-Thompson}, Z.~K., {et~al.} 2015, \apj,
  809, 77

\bibitem[{{Tingley}(2004)}]{2004A&A...425.1125T}
{Tingley}, B. 2004, \aap, 425, 1125

\bibitem[{{Torres} {et~al.}(2010){Torres}, {Bakos}, {Hartman}, {Kov{\'a}cs},
  {Noyes}, {Latham}, {Fischer}, {Johnson}, {Marcy}, {Howard}, {Sasselov},
  {Kipping}, {Sip{\H o}cz}, {Stefanik}, {Esquerdo}, {Everett}, {L{\'a}z{\'a}r},
  {Papp}, \& {S{\'a}ri}}]{2010ApJ...715..458T}
{Torres}, G., {Bakos}, G.~{\'A}., {Hartman}, J., {et~al.} 2010, \apj, 715, 458

\bibitem[{{Winn} {et~al.}(2008){Winn}, {Holman}, {Torres}, {McCullough},
  {Johns-Krull}, {Latham}, {Shporer}, {Mazeh}, {Garcia-Melendo}, {Foote},
  {Esquerdo}, \& {Everett}}]{2008ApJ...683.1076W}
{Winn}, J.~N., {Holman}, M.~J., {Torres}, G., {et~al.} 2008, \apj, 683, 1076

\bibitem[{{Winn} {et~al.}(2011){Winn}, {Howard}, {Johnson}, {Marcy},
  {Isaacson}, {Shporer}, {Bakos}, {Hartman}, {Holman}, {Albrecht}, {Crepp}, \&
  {Morton}}]{2011AJ....141...63W}
{Winn}, J.~N., {Howard}, A.~W., {Johnson}, J.~A., {et~al.} 2011, \aj, 141, 63

\bibitem[{{Young}(1967)}]{1967AJ.....72..747Y}
{Young}, A.~T. 1967, \aj, 72, 747

\end{thebibliography}

\end{document}